\documentclass[pdflatex, sn-vancouver, sort&compress]{sn-jnl}
\usepackage{gobble}
\usepackage{amsmath,amsfonts}
\usepackage{algorithmic}
\usepackage{array}
\usepackage{textcomp}
\usepackage{stfloats}
\usepackage{url}
\usepackage{verbatim}
\usepackage{graphicx}
\def\BibTeX{{\rm B\kern-.05em{\sc i\kern-.025em b}\kern-.08em
    T\kern-.1667em\lower.7ex\hbox{E}\kern-.125emX}}
\usepackage{balance}
\usepackage{manyfoot}
\usepackage{tabularx}
\usepackage{subcaption}
\usepackage{multicol}
\usepackage{wrapfig}

\usepackage{hyperref}  
\hypersetup{colorlinks=true,urlcolor=blue,citecolor=blue,linkcolor=blue}   
\usepackage[T1]{fontenc}	
\usepackage[latin9]{inputenc}	
\usepackage{geometry}		
\usepackage{graphicx}
\usepackage[above,below]{placeins}	
\usepackage{times}
\usepackage{multirow}
\usepackage[dvipsnames]{xcolor}
\usepackage{titlesec}
\usepackage{ifthen}
\usepackage{makecell}
\setcitestyle{numbers}
\setcitestyle{square}
\setcitestyle{comma}
\setcitestyle{sort&compress}

\newcommand{\fracroot}[2]{\ifthenelse{#1=1}{\frac{1}{\sqrt{#2}}}{\sqrt{\frac{#1}{#2}}}}

\usepackage{makecell}

\newif\ifshowtimestamp
\showtimestampfalse

\newlength{\clmax}
\newlength{\almax}
\def\dag{$^\dagger$}

\begin{document}
\title[Introductory QISE coursework at US institutions: Content coverage]{Introductory quantum information science coursework at US institutions: Content coverage}
\author*[1]{\fnm{Josephine C.} \sur{Meyer}}\email{josephine.meyer@colorado.edu}
\author[2]{\fnm{Gina} \sur{Passante}}\email{gpassante@fullerton.edu}
\author[1]{\fnm{Steven J.} \sur{Pollock}}\email{steven.pollock@colorado.edu}
\author[1]{\fnm{Bethany R.} \sur{Wilcox}}\email{bethany.wilcox@colorado.edu}
\affil*[1]{\orgdiv{Dept.\ of Physics}, \orgname{University of Colorado Boulder}, \orgaddress{\city{Boulder}, \state{CO}, \postcode{80309}, \country{USA}}}
\affil[2]{\orgdiv{Dept.\ of Physics}, \orgname{California State University Fullerton}, \orgaddress{\city{Fullerton}, \state{CA}, \postcode{92831}, \country{USA}}}

\abstract{
Despite rapid growth of quantum information science and engineering (QIS/QISE) workforce development initiatives, perceived lack of agreement among faculty on core content has made prior research-based curriculum and assessment development initiatives difficult to scale. To identify areas if consensus on content coverage, we report findings from a survey of N=63 instructors teaching introductory QISE courses at US institutions of higher learning. We identify a subset of content items common across a large fraction ($\ge$ 80\%) of introductory QISE courses that are potentially amenable to research-based curriculum development, with an emphasis on foundational skills in mathematics, physics, and engineering. As a further guide for curriculum development, we also examine differences in content coverage by level (undergraduate/graduate) and discipline. Finally, we briefly discuss the implications of our findings for the development of a research-based QISE assessment at the postsecondary level.
}


\maketitle

\section{Introduction and motivation}

The long-theorized Second Quantum Revolution \cite{Dowling:2003} is upon us, and educational initiatives in quantum information science and engineering (QIS/QISE\footnote{Quantum information science is increasingly referred to as ``quantum information science and engineering'' (QISE) to emphasize engineering's key role in this interdisciplinary field. We use the full acronym QISE here; our survey used the older but essentially interchangeable acronym QIS.}) are growing explosively at U.S.\ institutions and worldwide \cite{Plunkett:2020,Cervantes:2021,Dzurak:2022,Asfaw:2022}. A major driver for this growth has been the U.S.\ National Quantum Initiative Act of 2018 \cite{NQIA:2018,Raymer:2019}
and similar initiatives worldwide (e.g.\ \cite{Knight:2019,Riedel:2019,Zhang:2019}).
Quantum researchers and policymakers alike have expressed the need for educational programs to promote a quantum-ready workforce \cite{Fox:2020,Perron:2021,Hasanovic:2022,Hughes:2022,Singh:2022} and quantum-literate society \cite{Nita:2021}. Experts have accordingly called for education researchers to be involved in curriculum design from the start \cite{Marrongelle:2020, Aiello:2021}.

Yet a key challenge limiting involvement of discipline-based education research (DBER) in the development of research-based QISE curricular materials has been a lack of consensus on the goals of QISE education, much less core content coverage. In our own prior work, faculty have expressed disagreement on foundational issues ranging from the utility of programming activities to the merits of covering famous quantum algorithms \cite{Meyer:2022PhysRev,Meyer:2023TPT}. Finding points of similarity across courses is thus especially important if DBER researchers are to be involved in QISE education at scale.

Existing work provides a high-level framework to begin identifying these similarities. As early as the 2000s, educators began to theorize that -- much as a successful computer scientist can go an entire career paying minimal heed to the chemistry of silicon or the low-level circuitry of individual 0s and 1s -- relatively little quantum mechanics is needed to teach students quantum computing, as long as the emphasis is restricted to understanding and programming a quantum computer (as opposed to constructing one) \cite{Mermin:2003,Grau:2004,Singh:2006}. Seegerer \textit{et al.} \cite{Seegerer:2021} used interviews with QISE experts to conceptualize the breadth of QISE education, identifying key themes such as superposition, entanglement, quantum gates, quantum circuits, and quantum algorithms that they expected to be shared across courses. We aim to create a complementary roadmap at the granularity of discrete skills and concepts that can readily be translated into curricular materials, assessment items, or other tangible DBER outputs. In other words, we seek to identify the subset of core content (e.g.\ individual gates, algorithms, or mathematical concepts) sufficiently universal that DBER interventions will not remain confined to any one institution. For more precise targeting, we also examine how this material varies across types of courses.

\subsection{Research questions}

Three of our research questions specifically pertain to curriculum development:
\begin{itemize}
\item \textbf{RQ \#1}: Which QISE topics are sufficiently universal across courses to be strong candidates for research-based curriculum development?
\item \textbf{RQ \#2}: Which topics, if any, are more likely to be covered at the undergraduate vs.\ graduate levels? In other words, are certain topics' inclusion or non-inclusion the hallmark of a graduate or undergraduate QISE course?
\item \textbf{RQ \#3}: How does content coverage of introductory QISE courses vary across the academic discipline in which the course is offered? In other words, are there specific topics whose inclusion or non-inclusion is strongly associated with a specific discipline?
\end{itemize}

A second motivation for this work (\textbf{RQ \#4}) was to identify a subset of content instructors deem appropriate for inclusion in a research-based QISE assessment -- a stricter criterion than simply whether material is covered in a course:
\begin{itemize}
\item \textbf{RQ \#4}: What subset of QISE topics is suitable for developing a research-based QISE assessment?
\end{itemize}
Those readers specifically interested in assessment should refer to Sec.~\ref{sec:assessment}. The remainder of the paper is written for QISE educators and education researchers more broadly.

\section{Methodology}

\subsection{Survey design}

Informed by our group's previous experience designing and distributing faculty surveys to inform curriculum development and assessment efforts \cite{Meyer:2022PhysRev,Rainey:2019}, we developed a survey instrument enabling faculty to classify specific QISE topics by the degree of content coverage in the course. The survey was distributed electronically via Qualtrics.

\begin{figure*}
    \centering
    \fbox{
    \includegraphics[width = \textwidth]{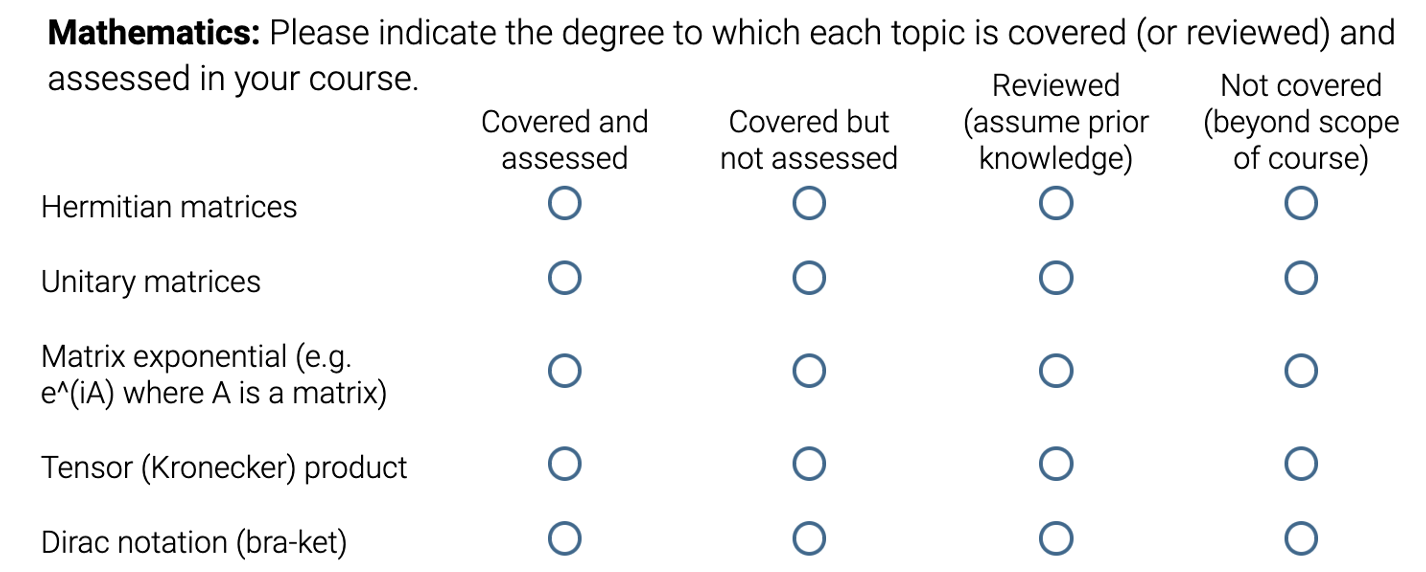}}
    \caption{A representative survey question. Screenshot shows top 5 items of a question on mathematical content. Faculty were asked to classify each content item by the degree of content coverage in their course.}
    \label{fig:qualtrics-screenshot}
\end{figure*}

We first compiled a list of QISE topics drawn from instructor responses to a prior open-ended survey \cite{Meyer:2022PhysRev}, contributed syllabi, and two commonly-assigned QISE textbooks (Refs.~\cite{Mermin:2007,Nielsen:2000}). Topics were revised, and a few added, based on feedback from QISE educators on our team. The list of topics was converted to survey questions in the manner of Fig.~\ref{fig:qualtrics-screenshot}. We also collected course background data (e.g.\ catalog listing) and optional respondent and institution demographics. 

Given the dual purposes of our study, we wished to distinguish between content coverage for curriculum development purposes and the stricter threshold of assessability. Accordingly, faculty were allowed to select 1 of 4 options for each content item: ``covered and assessed,'' ``covered but not assessed,'' ``reviewed (assume prior knowledge),'' and ``not covered (beyond scope of course).''
The survey instructions included the following definitions:

\begin{itemize}
\item \textbf{``Covered and assessed''} means the material is taught in your course and is fair game for the final exam or similar cumulative assessment. (If your course doesn't include a final exam, think about what it would include hypothetically.)
\item \textbf{``Covered but not assessed''} means you discuss the topic in class but it would not be appropriate for a final exam. This might include material that is discussed briefly or hand-wavily, or that is tangential to the primary goals of the course.
\end{itemize}

\subsection{Survey distribution}

Links to the survey were distributed over email to 449 faculty identified as teaching possible or probable QISE courses at U.S.\ institutions in the 2019-20, 2020-21, 2021-22, or 2022-23 academic years. Such QISE courses were identified via a thorough search of the course catalogs and associated databases of 475 institutions in fall 2022 \cite{Meyer:2023QST}.
The 475 institutions searched were selected based on quantity of degrees granted in QISE-adjacent fields. Institutions were included provided that they met one of the following criteria for at least one QISE-related field (physics, computer science, or electrical and/or computer engineering) in 2021:
\begin{itemize}
    \item Top 100 bachelor's degree-granting program in field, and/or awarded 50 or more bachelor's degrees in field
    \item Top 50 master's degree-granting program in field, and/or awarded 15 or more master's degrees in field
    \item Top 50 Ph.D.-granting program in field, and/or awarded 10 or more research-based Ph.D.'s in field
\end{itemize}

Email addresses were located for each faculty member identified as an instructor of one of the potential QISE courses identified from this search wherever possible. Email contacts do not necessarily correspond 1:1 with courses since a small number of instructors taught more than one course while a sizable number of courses were taught by multiple instructors. A small number of instructors (4) were added to the list because they had previously been identified as teaching QISE content via a similar but more limited study of course catalogs from 2019-20 \cite{Cervantes:2021}. Email recipients were encouraged to forward the survey to any other faculty teaching QISE courses they were aware of. The survey was open for approximately 6 weeks in September-October 2022; two follow-up reminder emails were sent during this period. 
 We received a total of $N=85$ substantially complete responses which we analyze in this article.

\subsection{Identifying introductory QISE courses}

For curriculum and assessment development purposes alike, a useful categorization is the ``introduction to QISE'' course, a subset of $N=63$ responses corresponding to introductory QISE courses at the undergraduate and/or graduate level. As in Ref.~\cite{Meyer:2022PhysRev},
we define introductory QISE courses as those that focus on theory of quantum information and/or quantum computing and that require no prerequisite QISE courses or domain-specific experience. Our definition excludes courses primarily focused on laboratory work or on materials/hardware development, though it includes courses that include QISE theory and another topic (e.g.\ quantum optics) as co-equal foci. Whether a course was classified as intro to QISE was determined primarily by self-classification; instructors could self-select one or more of the following:

\begin{itemize}
    \item Introduction to QISE/quantum computing (primarily focused on theory and/or programming)
    \item Introduction to QISE/quantum computing + an additional topic (e.g.\ quantum information + quantum optics)
    \item Hardware/materials course primarily focused on device fabrication or quantum hardware
    \item Beyond introductory QISE course (students expected to have already taken 1 or more dedicated QISE courses and/or have prior familiarity from research)
    \item Traditional quantum mechanics course, some QISE content
    \item Traditional computer science course, some QISE content
    \item Lab course or practicum
    \item Other/explain (free response)
\end{itemize}

The categories listed above were developed from responses to the 2021 survey \cite{Meyer:2022PhysRev} as well as analysis of course descriptions found via the catalog search. If an instructor selected either or both of the first two options (and no other option), the course was automatically coded as introduction to QISE. Our team manually coded courses for which the instructor selected multiple options and/or ``other/explain.'' Example ``other/explain'' responses classified as intro to QISE:

\begin{itemize}
    \item \textit{``Half quantum software and half hardware after intro to QM [quantum mechanics]''}
    \item \textit{``Graduate introduction to quantum information and computing requiring graduate QM but no previous QI[SE] experience''}
\end{itemize}
Example responses not coded as introduction to QISE:

\begin{itemize}
    \item \textit{``Mix of intro QM, quantum chemistry, DFT, supercomputing, and QIS[E]/quantum computing''} (too broad)
    \item \textit{``Studying cryptosystems which are/aren't secure in a quantum computing environment''} (advanced subtopic)
\end{itemize}

We focus on the $N=63$ courses coded by our team as introduction to QISE for the remainder of the paper.

\begin{figure}
    \centering
    \includegraphics[scale = 0.6]{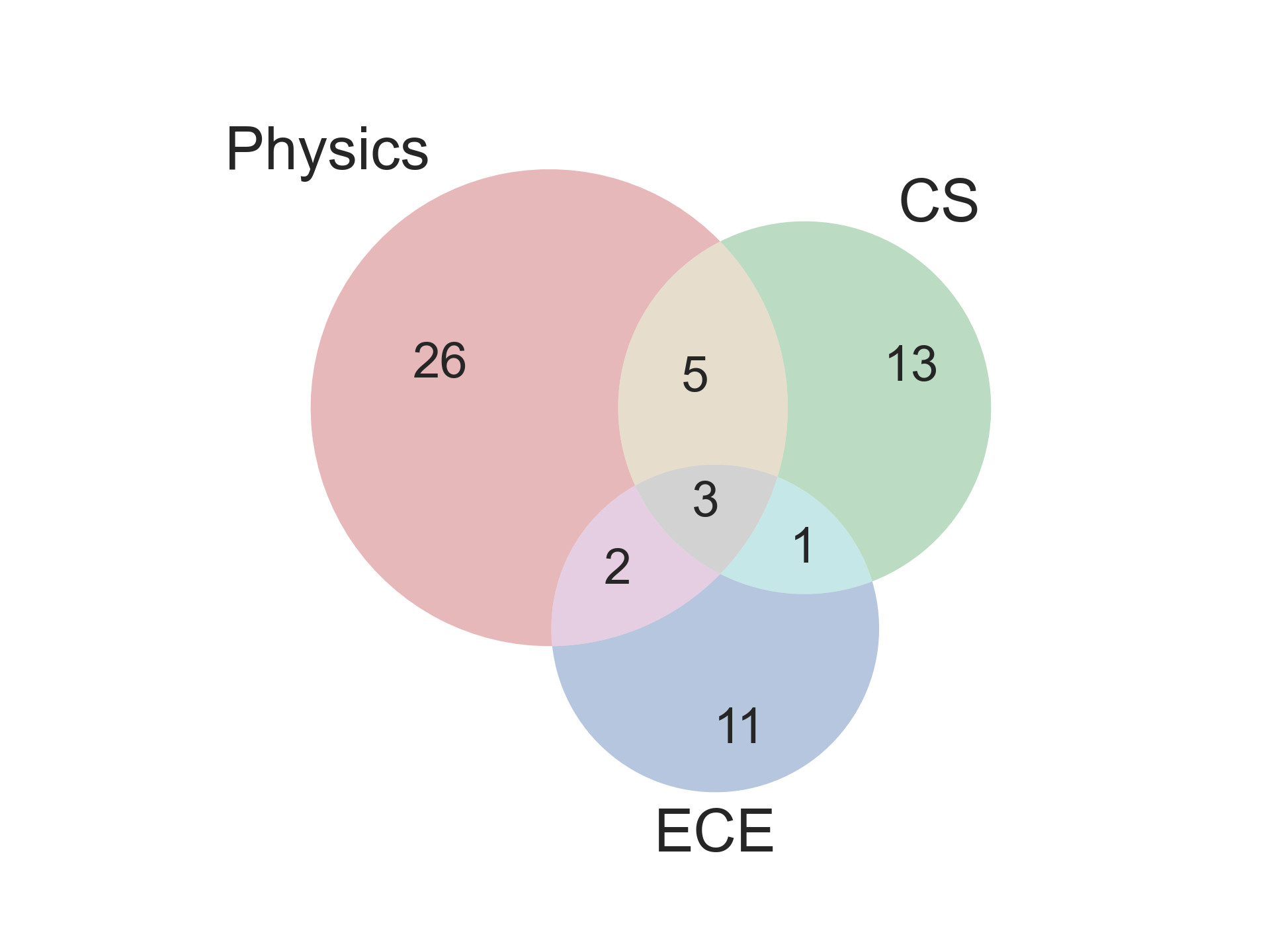}
    \caption{Catalog listings by discipline, including cross-listings, for each of the $N=63$ introductory QISE courses we discuss in this article. Note that $N=2$ courses are listed in none of these primary disciplines.}
    \label{fig:crosslistings}
\end{figure}

\subsection{Limitations}
\label{sec:limitations}

While our survey results represent the most diverse cross-section of QISE coursework in the US we are aware of in the literature, our survey results should be considered in light of a few key structural limitations. While we attempted to devise as complete a list of QISE courses as practically feasible when generating our instructor email list, assumptions made in the identification of courses may bias results in favor of instructors at certain institutions. Our dataset is restricted to 4-year institutions and does not reflect the experiences of instructors at 2-year colleges and pre-college programs, an audience that recent work argues has historically been underrepresented in the deployment of QISE educational initiatives \cite{Perron:2021}. 
(Of course, some instructors not on our email list may have received the survey forwarded from colleagues.)

Additionally -- reflecting the lack of diversity in QISE education -- our survey respondents were disproportionately white/Caucasian (82\% of those choosing to respond) and male (81\%). Respondents from large 4-year universities were also overrepresented in the dataset, consistent with prior work \cite{Cervantes:2021}: of the $N=59$ courses for which instructors submitted data on the institution where the course was taught, 43 (73\%) hailed from R1 research institutions and only $N=8$ (14\%) from non-doctoral-granting institutions. Approximately 75\% of courses were from publicly-funded institutions.

Finally, results may have been impacted by a pair of survey design anomalies. First, a Qualtrics glitch caused submissions for multiple courses in the same survey session to overwrite responses for prior courses, risking data loss. 
The survey was briefly re-opened in November 2022 to enable affected respondents to submit missing data; only one such response was submitted during this window. 
Since we expected a large majority of faculty to submit only a single response, 
we believe the glitch's effect on our data was small.

A second anomaly affected mathematical subtopics. The survey used Qualtrics display logic to automatically skip subtopic questions if the instructor selected ``covered minimally or not at all'' for the associated umbrella topic. However, the umbrella topic ``Mathematical Foundations of QIS'' appears to have been interpreted by some faculty as corresponding to a much greater mathematical sophistication than the research team anticipated. \footnote{For example, all 5 responses that selected ``cover minimally or not at all'' for mathematical foundations also selected ``covered and assessed'' for specific algorithms or communication protocols that would be extremely difficult to teach without linear algebra and bra-ket notation. Moreover, 2 of the 3 responses that selected ``covered but not assessed'' for ``mathematical foundations of QIS'' then marked ``covered and assessed'' for a large proportion of the math subtopics.} As designed, the survey automatically skipped all math subtopics for 5 instructors who selected ``covered minimally or not at all'' for ``mathematical foundations of QISE''; however, these 5 faculty's responses elsewhere in the survey are inconsistent with a math-free course.  For this reason, math subtopics are reported as percentages only out of the $N=58$ courses whose instructors were explicitly shown these subtopics.

\begin{table*}

\setlength{\tabcolsep}{1pt}
    \centering
    \begin{tabular}[t]{l  @{\hskip-.5cm} c}
        \hline \hline
         \thead{Topic} & \thead{\% Covered}\\
        \Xhline{2\arrayrulewidth}\vspace{-6pt}\\ 
        \textbf{Qubits}* & \textbf{100\%} \\\vspace{-6pt}\\
        \textbf{Entanglement}* & \textbf{100\%} \\
        \vspace{-6pt}\\
        \textbf{Superposition}* & \textbf{98\%} \\
        \vspace{-6pt}\\
        \textbf{Quantum gates}* & \textbf{98\%}\\
         \ \ CNOT* & \textbf{98} \\ 
         \ \ Hadamard (H)* & \textbf{98} \\
         \ \ Identity (I)* & \textbf{98} \\
         \ \ Pauli X* & \textbf{98} \\
         \ \ Pauli Z* & \textbf{98} \\
         \ \ Pauli Y* & \textbf{97}\\
         \ \ Phase gate (S, T) & \textbf{92} \\
         \ \ SWAP & \textbf{89} \\
         \ \ CZ & \textbf{89}  \\
         \ \ Universality of gates & \textbf{89} \\
         \ \ Toffoli (CCNOT) & \textbf{84} \\
         \ \ Other 2+ qubit gate(s) & \textit{68} \\ \vspace{-6pt}\\
        \textbf{Quantum algorithms}* & \textbf{98\%}\\
         \ \ Deutsch/Deutsch-Jozsa* & \textbf{90} \\ 
         \ \ Grover search algorithm & \textbf{87} \\
         \ \ Quantum Fourier transform & \textbf{87} \\
         \ \ Shor's factoring algorithm & \textit{79} \\
         \ \ Simon's problem & \textit{62} \\
         \ \ Bernstein-Vazirani & \textit{56} \\
         \ \ Phase estimation subroutine & \textit{52} \\
         \ \ Quantum machine learning & \textit{17} \\
         \ \ Other algorithm(s) & \textit{13}\\

         \vspace{-6pt}\\

        \textbf{Quantum measurement}* & \textbf{97\%} \\ \vspace{-6pt}\\

        \textbf{Quantum circuit} & \textbf{97\%} \vspace{-3pt}\\
        \ \ \textbf{diagrams}* & \vspace{-6pt}\\

        \end{tabular} 
        \vspace{-24pt}
        \begin{tabular}[t]{l @{\hskip-.42cm} c}
        \hline \hline
         \thead{Topic} & \thead{\% Covered} \\
        \Xhline{2\arrayrulewidth}\vspace{-6pt}\\ 

       \textbf{Math foundations QISE}*\dag & \textbf{(92\%)} \\
         \ \ Dirac notation (bra-ket)* & \textbf{100} \\ 
         \ \ Complex numbers* & \textbf{100} \\
         \ \ Unitary matrices* & \textbf{100} \\
         \ \ Inner product* & \textbf{98} \\
         \ \ Dim.\ of Hilbert space* & \textbf{97} \\
         \ \ Vector spaces, finite dim.* & \textbf{93} \\
         \ \ Tensor/Kronecker product* & \textbf{93} \\
         \ \ Hermitian matrices & \textbf{93} \\
         \ \ Eigenvalues/eigenvectors* & \textbf{91} \\
         \ \ Outer product & \textbf{86} \\
         \ \ Matrix exponential & \textit{79} \\
         \ \ Commutators & \textit{66} \\
         \ \ Vector spaces, infinite dim. & \textit{43} \\
         \ \ Number theory & \textit{40} \\
         \ \ Combinatorics & \textit{31}\\ \vspace{-6pt} \\

         \textbf{Quantum comm./cryptog.}* & \textbf{87\%}\\ 

         \ \ Quantum teleportation & \textbf{81} \\ 
         \ \ Quantum key dist.\ (BB84) & \textit{70} \\
         \ \ Superdense coding & \textit{60} \\
         \ \ Quantum key dist.\ (E91) & \textit{30} \\
         \ \ Quantum key dist.\ (B92) & \textit{27} \\
         \ \ Quantum key dist.\ (other) & \textit{3}\\ \vspace{-6pt}\\

        \textbf{Bloch sphere} & \textbf{84\%} \\ \vspace{-6pt}\\

        \end{tabular} 
        \vspace{-24pt}
        \begin{tabular}[t]{l @{\hskip-.5cm} c}
        \hline \hline
         \thead{Topic} & \thead{\% Covered} \\
        \Xhline{2\arrayrulewidth}\vspace{-6pt}\\ 

        \textbf{Physical implementations} & \textit{71\%} \\
        \ \ Superconducting qubits & \textit{57} \\ 
         \ \ Trapped ions & \textit{57}\\
         \ \ Photonics & \textit{44}\\ 
         \ \ Neutral atoms & \textit{25} \\
         \ \ Nitrogen vacancy centers & \textit{24}\\
         \ \ Other implementation(s) & \textit{11} \\ \vspace{-6pt}\\

         \textbf{QISE theory (uncategorized)} & -- \vspace{2pt} \\
         \ \ No cloning theorem & \textbf{95}  \\
         \ \ EPR paradox & \textbf{81}  \\
         \ \ Bell inequalities & \textit{75} \\
         \ \ Decoherence, noise channels & \textit{73} \\
         \ \ Quantum error corr.\ codes & \textit{71} \\
         \ \ Heisenberg uncertainty & \textit{65}  \\
         \ \ Open quantum systems & \textit{56} \\
         \ \ Complexity classes & \textit{49} \\ \vspace{-6pt}\\

    \textbf{Advanced topics} &--  \vspace{2pt} \\
         \ \ Density matrix/mixed states & \textit{68} \\
         \ \ Coherence time (T1, T2) & \textit{38} \\
         \ \ State fidelity & \textit{37} \\
         \ \ Quantifying entanglement & \textit{35} \\
         \ \ Entropy (von Neumann) & \textit{33} \\
         \ \ Entropy (Shannon) & \textit{33} \\
         \ \ Adiabatic QC & \textit{29} \\
         \ \ Measurement-based QC & \textit{24} \\
         \ \ Quantum sensing/metrology & \textit{22} \\
        \ \ Quantum optics & \textit{19} \\
        \ \ Quantum compilers & \textit{17} \\
        \ \ Topological QC & \textit{14} \\
        \ \ Post-quantum classical & \textit{11} \vspace{-3pt} \\
        \ \ \ \ cryptography & \\
    \end{tabular}

    \vspace{1.8cm}
    \hrulefill
    \vspace{0.2cm}

    \caption{Full survey results for introductory QISE courses at the postsecondary level ($N=63$ courses). Bolded percentages reflect those topics covered in $\ge80\%$ of courses. $\ast$Topics also meeting the stricter threshold of being \textit{assessed} in $\ge80\%$ of courses (see Sec.~\ref{sec:assessment}). \dag Percentages for mathematics subtopics are reported only for the $N=58$ responses shown full list of subtopics (see Sec.~\ref{sec:limitations}). }
    \label{tab:full-results}
\end{table*}

\section{Analysis and key findings}

For \textbf{RQ \#1-3}, we collapse the categories of ``covered and assessed'' and ``covered but not assessed'' as both levels of coverage are generally suitable for curricular materials development and for comparative analysis of curricula. We also include ``reviewed (assume prior knowledge)'' in the collapsed category because QISE instructors often have to spend substantial time on foundational mathematics and physics content even if it has been ostensibly seen in prerequisite courses \cite{Meyer:2022PhysRev}, and reviewed material often must nevertheless be (re)learned independently outside of lecture. For \textbf{RQ \#4} we collapse only ``covered and assessed'' and ``reviewed (assume prior knowledge),'' as discussed in Sec.~\ref{sec:assessment-methods}.

A complete list of topics and subtopics, ordered by the percentage of courses covering them, is provided in Table~\ref{tab:full-results}. We discuss specific findings below by research question.

\subsection{\textbf{RQ \#1}: Core curriculum for curriculum development}
\label{sec:high-level}

We identify a core set of topics that appear to be shared across a large majority ($\ge 80\%$) of introductory QISE courses and that represent foundational skills and concepts that can be effectively targeted for curriculum development (listed here, and bolded in Table~\ref{tab:full-results}):

\begin{itemize}
    \item \textbf{Qubits}
    \item \textbf{Entanglement}
    \item \textbf{Superposition}
    \item \textbf{Quantum gates} (including most well-known gates and the concept of quantum gate universality)
    \item \textbf{Quantum algorithms} (specifically Deutsch-Jozsa, Grover's search, and the Quantum Fourier Transform)
    \item \textbf{Quantum measurement}
    \item \textbf{Quantum circuit diagrams}
    \item \textbf{Quantum communication and cryptography} (specifically quantum teleportation)
    \item \textbf{Bloch sphere}
    \item \textbf{Certain mathematical skills}: Dirac notation, complex numbers, finite-dimensional Hilbert spaces and their dimension, unitary and Hermitian matrices, inner/outer/tensor products, and eigenvalues/eigenvectors
    \item \textbf{Miscellaneous QISE theory topics}: no cloning theorem, EPR paradox
\end{itemize}

Note that the 80\% threshold was chosen empirically (see Table \ref{tab:full-results} exhibiting a visible gap in reported percentages near this threshold) as a first-order partition of the surveyed content for curriculum development. Readers are encouraged to apply a higher or lower threshold as appropriate to their work and should interpret the percentages reported in Table~\ref{tab:full-results} accordingly.

\subsection{\textbf{RQ \#2}: Comparing undergraduate and graduate courses}
\label{sec:levels-comparison}

Few meaningful differences were observed between undergraduate and graduate courses in terms of content coverage.
Only for the three advanced  topics shown in Table~\ref{tab:gradundergrad} were statistically significant differences detected between graduate and undergraduate courses (first-order Rao-Scott\footnote{Rao-Scott modified $\chi^2$ \cite{Rao:1981} is an extension of Pearson $\chi^2$ to multiple-response datasets, necessitated by the relatively large number of hybrid undergraduate/graduate courses (22 out of $N=63$ courses). See Ref.~\cite{Decady:2000} for an accessible breakdown of the first-order approximation applied here.
} modified $\chi^2$, $p_{adj} < 0.05$ with Hochberg-Bonferroni adjustment\footnote{The Hochberg-Bonferroni procedure is designed to compensate for the risk of incorrectly rejecting the null hypothesis (Type I error) given repeated statistical tests while retaining greater statistical power than the conservative Holm-Bonferroni adjustment \cite{Hochberg:1988}.
Holm-Bonferroni is designed to strictly control Type I error (false positives) but can lead to high rates of Type II error (false negatives) for datasets with few significant variables. For our exploratory purposes, we judge Type II error of comparable concern to Type I error so a less strict, though still conservative, adjustment is warranted.} for multiple statistical tests). Moreover, these three advanced topics are closely related: von Neumann entropy is a basis-independent quantum extension of Shannon entropy, and mixed states and the associated mathematical tool of density matrices enable the entropy of a quantum state to be discussed quantitatively. In short, it appears that a primary distinction between undergraduate and graduate introductory QISE courses is that graduate courses are more likely to introduce the concept of mixed states and the associated information-theoretic concept of entropy.
Otherwise, the percentage of undergraduate and graduate courses covering each topic were within an absolute difference of 12\%, giving us confidence that our findings in Sec.~\ref{sec:high-level} generalize across both graduate and undergraduate levels.

\begin{table*}
    \setlength\tabcolsep{6 pt}
    \centering
    \begin{tabular}{l c c c c }
        \hline
        \hline
        \thead{Topic} & \thead{\%Undergraduate} & \thead{\% Graduate} & \thead{$p$} &\thead{$p_{adj}$} \\
        \Xhline{2\arrayrulewidth} \vspace{-6pt}\\
        Entropy (Shannon) & 26\% & 49\% & $p < 0.001$ & $p_{adj} < 0.001$\\
        
        Density matrices/mixed states & 63\% & 82\% & $p = 0.002$ & $p_{adj} = 0.004$\\

        Entropy (von Neumann) & 26\% & 44\% & $p = 0.011$ & $p_{adj} = 0.033$\\ 
        \hline

    \end{tabular}
    \caption{Topics for which a statistically significant difference ($p_{adj} < 0.05$) was observed in coverage between undergraduate and graduate introduction to QISE courses.  $N=63$ courses: $N_U = 46$ listed at the undergraduate level, $N_G = 39$ at the graduate level (22 listed as both).}
    \label{tab:gradundergrad}
\end{table*}

\subsection{\textbf{RQ \#3}: Comparison of courses across disciplines}
\label{sec:disciplines}

\begin{table*}
    \setlength\tabcolsep{5 pt}
    \centering
    \begin{tabular}{l c c c c}
        \hline
        \hline
        \thead{Topic} & \thead{\%CS} & \thead{\%NotCS} & \thead{$p$} &\thead{$p_{adj}$} \\
        \Xhline{2\arrayrulewidth} \vspace{-6pt}\\
        Commutators* & 30\% & 84\% & $p < 0.001$ & $p_{adj} < 0.001$\\
        
        Quantum sensing \& metrology & 0\% & 34\% & $p = 0.001$ & $p_{adj} = 0.002$\\

        Density matrices/mixed states & 41\% & 83\% & $p = 0.001$ & $p_{adj} = 0.004$\\ 

        Outer product* & 65\% & 97\% & $p = 0.002$ & $p_{adj} = 0.006$ \\ 

        Shor's factoring algorithm\ & 100\% & 68\% & $p = 0.002$ & $p_{adj} = 0.012$ \\ 
        \hline
        \thead{Topic} & \thead{\%Physics} & \thead{\%NotPhysics} & \thead{$p$} &\thead{$p_{adj}$} \\
        \Xhline{2\arrayrulewidth} \vspace{-6pt}\\

        Deutsch (or Deutsch-Jozsa) algorithm & 100\% & 78\% & $p = 0.004$ & $p_{adj} = 0.026$ \\ 

        Quantum error correcting codes\  & 86\% & 52\% & $p = 0.005$ & $p_{adj} = 0.032$ \\

        \hline

    \end{tabular}
    \caption{Topics for which a statistically significant difference ($p_{adj} < 0.05$) was observed in coverage across disciplines. $N=63$ (*$N=58$) courses: $N_{Phys} = 36$ (32) listed in physics or engineering physics, $N_{CS} = 22$ (20) listed in computer science, $N_{ECE} = 17$ (17) listed in ECE.}
    \label{tab:listing}
\end{table*}

As a proxy for intended academic audience, we classified courses by catalog listing\footnote{We recognize that per Ref.~\cite{Meyer:2022PhysRev}, course catalog listing does not always correlate well with the actual breakdown of the students in a QISE course; however, we take catalog listing as a useful indicator of the audience(s) most explicitly being targeted.}: 
physics, computer science, electrical and/or computer engineering (ECE), or other.\footnote{Physics includes engineering physics. ECE includes one transdisciplinary program encompassing several engineering subfields, electrical among them.} We observed a few notable differences in content coverage across disciplines amid a general trend of cross-disciplinary uniformity. Table~\ref{tab:listing} shows topics where statistically-significant differences were detected between courses inside and outside a particular discipline ($p_{adj} < 0.05$, Fisher's exact\footnote{Unlike Sec.~\ref{sec:levels-comparison}, we apply Fisher's exact test here because because a large proportion of contingency tables violate the assumptions of a $\chi^2$ test due to small expected counts. (On the other hand, we know of no accessible procedure for expanding Fisher's exact to multiple-response datasets, motivating the use of Rao-Scott modified $\chi^2$ in Sec.~\ref{sec:levels-comparison} where expected counts were sufficiently large to justify using a non-exact test.)}
with Hochberg-Bonferroni adjustment). No significant differences were found between ECE and non-ECE courses, though this may be an artifact of small sample size ($N_{\text{ECE}}=17$).

Computer science courses disproportionately taught Shor's factoring algorithm (100\% of CS-listed courses, 68\% of non-CS courses). We hypothesize that CS instructors may place greater value on the explicit connections between quantum computing and integer factoring (which is a problem of specific importance to cybersecurity) than physicists and electrical engineers.
Indeed, no significant difference ($p = .24$, Fisher's exact) is observed in CS vs.\ non-CS coverage of the quantum Fourier transform -- the core quantum subroutine of Shor's algorithm -- suggesting that the underlying distinction is in treatment of the classical subroutines linking the quantum Fourier transform to factoring. Additionally, computer science courses were disproportionately likely \textit{not} to teach commutators, outer products, density matrices/mixed states, and quantum sensing/metrology -- all topics of great interest to quantum information theory and the physical construction and characterization of quantum technologies but with fewer direct connections to quantum algorithms.

Finally, two topics -- Deutsch's algorithm and quantum error correcting codes -- were statistically more likely to be taught in physics than in non-physics courses. One explanation for the former may be that Deutsch's algorithm is arguably the most accessible quantum algorithm to demonstrate meaningful quantum advantage (albeit on a problem of purely theoretical interest) and thus might be the go-to algorithm for instructors who just wish to demonstrate a single algorithm to show how quantum advantage occurs theoretically.

\subsection{\textbf{RQ \#4}: Implications for research-based QISE assessment}
\label{sec:assessment}
\label{sec:assessment-methods}

Research-based conceptual assessments such as the Force Concept Inventory \cite{Hestenes:1992} have historically been powerful DBER tools, with benefits ranging from helping educators improve their teaching methods via reliable comparisons across instructors and institutions to validating research-based curricular materials \cite{Wilcox:2015Review}.
In 2021, 34 prominent leaders in quantum research and education called for emerging QISE education programs to commit to utilizing research-based assessments from the DBER community from the beginning so as to identify and promulgate best practices \cite{Aiello:2021}.
While a number of research-based quantum mechanics assessments have been developed \cite{Cataloglu:2002,Goldhaber:2009,Wuttiprom:2009,McKagan:2010,Singh:2010,Sadaghiani:2014,Marshman:2019},
it is our group's experience that none appears particularly well-aligned with the goals and content emphases of a typical introductory QISE course \cite{Madsen:2017}. Since a new assessment instrument targeting postsecondary introductory QISE courses may be warranted, we find it useful to compile a more specific list comprising those subset of topics that instructors \textit{assess} in their courses.

For assessment purposes, we collapse only the categories of ``covered and assessed'' and ``reviewed (assume prior knowledge)'' in order to preserve the covered/assessed distinction. Those topics that met an 80\% \textit{assessability} threshold are indicated in Table~\ref{tab:full-results} by an asterisk.\footnote{The 80\% assessability threshold, like the curriculum development threshold, was chosen empirically based on visual analysis of the data.} While instructor reports of what material is assessed in there classes are a useful starting point, we emphasize that not all topics included on this list will ultimately be appropriate for inclusion on a research-based QISE assessment. Formulating actionable assessment objectives is itself an involved, iterative procedure \cite{Vignal:2022} beyond the scope of this paper that we anticipate will winnow our list of topics significantly as some topics (e.g.\ quantum gates) are more readily translated to assessment objectives than others (e.g.\ Deutsch's algorithm).
\section{Discussion and conclusions}
\label{sec:conclusions}

A focus-group study of experts by Seegerer \textit{et al.} \cite{Seegerer:2021} identified five core ideas in quantum computing education: superposition, entanglement, quantum computers, quantum algorithms, and quantum cryptography. Two other studies have used comparable methodologies to come to similar conclusions \cite{Alpert:2020, Gerke:2022}; of course, each study's respective lists of specific ideas and competencies varies in terms of length and granularity. Our study provides a complementary perspective to this growing body of literature, asking what material is actually covered in real courses (as opposed to expert opinions of what ought to be included).

In fact, we find that Seegerer \textit{et al.}'s core ideas correspond quite strongly to the umbrella topics we find are covered in $\ge 80\%$ of the intro QISE courses we surveyed: Superposition, entanglement, quantum algorithms, and quantum communication. Our presumed umbrella topics of qubits, quantum gates, and quantum measurement together constitute the fundamental building blocks of Seegerer \textit{et al.}'s concept of the \textit{quantum computer}, while our umbrella topics of mathematical foundations of QISE, quantum circuit diagrams, and the Bloch sphere represent specific representations for conceptualizing the above ideas. Meanwhile, physical implementations of quantum computing -- our only presumed umbrella topic that does not map neatly to any of Seegerer \textit {et al.}'s categories -- was covered in a lower percentage of courses (71\%) than any comparable topic. Seegerer \textit{et al.} present their findings as investigating QISE education from a computer science perspective; here we show that their findings can be generalized to QISE education more broadly.

We also see evidence that each discipline -- physics, computer science, and ECE -- has a unique perspective on QISE education. In fact, the disproportionate number of CS-specific differences we observe seem to confirm Seegerer \textit{et al.}'s contention that a CS-specific perspective on QISE education may exist. (Perhaps the CS perspective is best identified by the focus on software and algorithms over hardware, in which case the absence of physical implementations from Seegerer \textit{et al.}'s five core ideas of QISE education is unsurprising). Nevertheless, on the whole, we find that  similarities in content converge across disciplines far outnumber differences. We see the convergence of these curricula across disciplines, along with the sizable number of cross-listed courses, as evidence that a core QISE curriculum is coalescing even in the face of the widespread variation in learning goals and instructor backgrounds we observed in 2021 \cite{Meyer:2022PhysRev}.

As outlined in Sec.~\ref{sec:high-level} and shown in Table~\ref{tab:full-results}, we find that certain topics are covered in a large majority ($\ge 80\%$) of introductory QISE courses and as such make ideal targets for developers of research-based curricular materials. A similar list is developed for assessment. While introductory QISE courses may vary in terms of what advanced topics may be covered, there remains a common set of foundational math and physics concepts that undergird these higher-level topics. In prior work \cite{Meyer:2022PhysRev}, tensor products and eigenvalues/eigenvectors emerged as particular sites of instructor-perceived difficulty for students, so these two topics may be ideal starting points for curriculum development. Alternatively, tutorials breaking down the most commonly taught quantum algorithms -- Deutsch-Jozsa, Grover's search, and the Quantum Fourier Transform -- into conceptually digestible pieces may also prove valuable since these algorithms require synthesizing many distinct and challenging concepts encountered throughout the course.

\subsection{Guidance and future work}

Over the past few years, DBER communities have begun heeding the community's calls for research-based curriculum development \cite{Marrongelle:2020,Aiello:2021}, creating QISE education materials for a variety of contexts (e.g. \cite{Salehi:2022}). However, such work has been relatively specialized, often focusing on teaching quantum computing to high school students \cite{
Angara:2022,Bondani:2022, Satanassi:2022} and developing simulation resources \cite{Kohnle:2017,DeVore:2020}. We urge the community to broaden our focus to span more areas of QISE theory, and especially to build on curriculum development projects such as QuSTEAM \cite{Porter:2022} and Qubit by Qubit \cite{QBQ} that center needs of historically underrepresented students. 
We likewise call on the community to place greater emphasis on fundamental research in QISE student reasoning (e.g.\ \cite{
Meyer:2022PERC,Kushimo:2022}), which can serve to scaffold future curriculum development initiatives.

Finally, we reiterate that the core content list we have developed here is just one intermediary step toward the development of research-based curricular materials or assessment items. Our future work will build upon this list of core content to formulate measurable learning outcomes for development and validation of curricular materials. (The Quantum Curriculum Transformation Framework \cite{Goorney:2023} provides a useful framework for subsequently translating these learning outcomes into effective curricula.) Separately, we have begun to use the list compiled in Sec.~\ref{sec:assessment} to develop objectives and draft assessment items for a research-based postsecondary QISE assessment,\footnote{While our work targets the introductory undergraduate/graduate audience, an EU-based collaboration is developing a parallel assessment instrument for high school students \cite{Faletic:2023}.} which based on the findings in Table~\ref{tab:full-results} will most likely emphasize quantum computing.

\backmatter

\bmhead{Availability of data and materials}
The datasets generated and/or analysed during the current study are not publicly available due to identifiable human subjects data, but can be made available (with necessary redactions) from the corresponding author on reasonable request.

\bmhead{Competing interests}
The authors declare that they have no competing interests.

\bmhead{Funding}
This work was supported by the University of Colorado Boulder Dept.\ of Physics, the NSF Graduate Research Fellowship Program, and NSF Grants No.'s 2012147 and 2011958.

\bmhead{Authors' contributions}
 JCM developed the first draft of the survey and performed the majority of the data collection, analysis, and writing with input and guidance from GP, SJP, and BRW. All authors contributed considerably to the design and content of the survey instrument as well as the formulation of research questions. All authors have read and approved the final manuscript.

\bmhead{Acknowledgments}
Special thanks to Beth Thacker for feedback on the survey design and Catherine Talbot for advice on statistical tests. We also thank our faculty respondents, both those who responded to this survey and those whose contributed syllabi in 2021 informed content of this survey.

\newpage

\bibliography{ref}

\begin{thebibliography}{10}
\providecommand{\doi}[1]{\url{https://doi.org/#1}}
\bibcommenthead

\bibitem[\protect\citeauthoryear{Dowling and Milburn}{{2003}}]{Dowling:2003}
Dowling JP, Milburn GJ.
\newblock {Quantum technology: The second quantum revolution}.
\newblock {Phil Trans Royal Soc A}. {2003};{361}({1809}):1655--1674.
\newblock \doi{10.1098/rsta.2003.1227}.

\bibitem[\protect\citeauthoryear{Plunkett et~al.}{{2020}}]{Plunkett:2020}
Plunkett T, Frantz TL, Khatri H, Rajendran P, Midha S.
\newblock {A survey of educational efforts to accelerate a growing quantum
  workforce}.
\newblock In: {Proc. 2020 IEEE Int. Conf. Quantum Comput. Eng.}; {2020}. p.
  {330--336}.
\newblock \doi{10.1109/QCE49297.2020.00048}.

\bibitem[\protect\citeauthoryear{Cervantes et~al.}{{2021}}]{Cervantes:2021}
Cervantes B, Passante G, Wilcox B, Pollock S.
\newblock {An overview of quantum information science courses at US
  institutions}.
\newblock In: {Proc. 2021 Phys. Educ. Res. Conf.}; {2021}. p. {93--98}.
\newblock \doi{10.1119/perc.2021.pr.Cervantes}.

\bibitem[\protect\citeauthoryear{Dzurak et~al.}{{2022}}]{Dzurak:2022}
Dzurak AS, Epps J, Laucht A, Malaney R, Morello A, Nurdin HI, et~al.
\newblock {Development of an undergraduate quantum engineering degree}.
\newblock {IEEE Trans Quantum Eng}. {2022};{3}:{6500110}.
\newblock \doi{10.1109/TQE.2022.3157338}.

\bibitem[\protect\citeauthoryear{Asfaw et~al.}{{2022}}]{Asfaw:2022}
Asfaw A, Blais A, Brown KR, Candelaria J, Cantwell C, Carr LD, et~al.
\newblock {Building a quantum engineering undergraduate program}.
\newblock {IEEE Trans Educ}. {2022};{65}({2}):{220--242}.
\newblock \doi{10.1109/TE.2022.3144943}.

\bibitem[\protect\citeauthoryear{{115th US Congress (2018)}}{}]{NQIA:2018}
{115th US Congress (2018)}.: {H.R.6227 National Quantum Initiative Act}.

\bibitem[\protect\citeauthoryear{Raymer and Monroe}{{2019}}]{Raymer:2019}
Raymer MG, Monroe C.
\newblock {The US national quantum initiative}.
\newblock {Quantum Sci Technol}. {2019};{4}:{020504}.
\newblock \doi{10.1088/2058-9565/ab0441}.

\bibitem[\protect\citeauthoryear{Knight and Walmsley}{{2019}}]{Knight:2019}
Knight P, Walmsley I.
\newblock {UK national quantum technology programme}.
\newblock {Quantum Sci Technol}. {2019};{4}:{040502}.
\newblock \doi{10.1088/2058-9565/ab4346}.

\bibitem[\protect\citeauthoryear{Riedel et~al.}{{2019}}]{Riedel:2019}
Riedel M, Kovacs M, Zoller P, Mlynek J, Calarco T.
\newblock {Europe's quantum flagship initiative}.
\newblock {Quantum Sci Technol}. {2019};{4}:{020501}.
\newblock \doi{10.1088/2058-9565/ab042d}.

\bibitem[\protect\citeauthoryear{Zhang et~al.}{{2019}}]{Zhang:2019}
Zhang Q, Xu F, Li L, Liu NL, Pan JW.
\newblock {Quantum information research in China}.
\newblock {Quantum Sci Technol}. {2019};{4}:{040503}.
\newblock \doi{10.1088/2058-9565/ab4bea}.

\bibitem[\protect\citeauthoryear{Fox et~al.}{{2020}}]{Fox:2020}
Fox MFJ, Zwickl BM, Lewandowski HJ.
\newblock {Preparing for the quantum revolution: What is the role of higher
  education?}
\newblock {Phys Rev Phys Educ Res}. {2020};{16}:{020131}.
\newblock \doi{10.1103/PhysRevPhysEducRes.16.020131}.

\bibitem[\protect\citeauthoryear{Perron et~al.}{{2021}}]{Perron:2021}
Perron JK, DeLeone C, Sharif S, Carter T, Grossman JM, Passante G, et~al.
\newblock {Quantum undergraduate education and scientific training}.
\newblock arXiv.
  {2021};{\href{https://arxiv.org/abs/2109.13850}{{2109.13850}}}.

\bibitem[\protect\citeauthoryear{Hasanovic et~al.}{{2022}}]{Hasanovic:2022}
Hasanovic M, Panayiotou CA, Silberman DM, Stimers P, Merzbacher CI.
\newblock {Quantum technician skills and competences for the emerging Quantum
  2.0 industry}.
\newblock {Opt Eng}. {2022};{61}:{081803}.
\newblock \doi{https://doi.org/10.1117/1.OE.61.8.081803}.

\bibitem[\protect\citeauthoryear{Hughes et~al.}{{2022}}]{Hughes:2022}
Hughes C, Finke D, German DA, Merzbacher C, Vora PM, Lewandowski HJ.
\newblock {Assessing the needs of the quantum industry}.
\newblock {IEEE Trans Educ}. {2022};{65}({4}):{592--601}.
\newblock \doi{doi: 10.1109/TE.2022.3153841}.

\bibitem[\protect\citeauthoryear{Singh et~al.}{{2022}}]{Singh:2022}
Singh C, Levy A, Levy J.
\newblock {Preparing precollege students for the Second Quantum Revolution with
  core concepts in quantum information science}.
\newblock {Phys Teach}. {2022};{60}({8}):{639--641}.
\newblock \doi{10.1119/5.0027661}.

\bibitem[\protect\citeauthoryear{Nita et~al.}{{2023}}]{Nita:2021}
Nita L, Smith LM, Chancellor N, Cramman H.
\newblock {The challenge and opportunities of quantum literacy for future
  education and transdisciplinary problem solving}.
\newblock {Res Sci Technol Ed}. {2023};41(2):564--580.
\newblock \doi{10.1080/02635143.2021.1920905}.

\bibitem[\protect\citeauthoryear{Marrongelle}{}]{Marrongelle:2020}
Marrongelle K.: {Dear colleague letter: Advancing quantum education and
  workforce development}.
\newblock NSF 21-033, 2020.

\bibitem[\protect\citeauthoryear{Aiello et~al.}{{2021}}]{Aiello:2021}
Aiello CD, Awschalom DD, Bernien H, Brower T, Brown KR, Brun TA, et~al.
\newblock {Achieving a quantum smart workforce}.
\newblock {Quantum Sci Technol}. {2021};{6}:{030501}.
\newblock \doi{10.1088/2058-9565/abfa64}.

\bibitem[\protect\citeauthoryear{Meyer et~al.}{{2022}}]{Meyer:2022PhysRev}
Meyer JC, Passante G, Pollock SJ, Wilcox BR.
\newblock {Today's interdisciplinary quantum information classroom: Themes from
  a survey of quantum information science instructors}.
\newblock {Phys Rev Phys Educ Res}. {2022};{18}:{010150}.
\newblock \doi{10.1103/PhysRevPhysEducRes.18.010150}.

\bibitem[\protect\citeauthoryear{Meyer et~al.}{{2023}}]{Meyer:2023TPT}
Meyer JC, Passante G, Pollock SJ, Wilcox BR.
\newblock {How media hype affects our physics teaching: A case study on quantum
  computing}.
\newblock {Phys Teach}. {2023};{61}(5):{339--342}.
\newblock \doi{10.1119/5.0117671}.

\bibitem[\protect\citeauthoryear{Mermin}{{2003}}]{Mermin:2003}
Mermin ND.
\newblock {From Cbits to Qbits: Teaching computer scientists quantum
  mechanics}.
\newblock {Am J Phys}. {2003};{71}:{23--30}.
\newblock \doi{10.1119/1.1522741}.

\bibitem[\protect\citeauthoryear{Grau}{{2004}}]{Grau:2004}
Grau BC.
\newblock {How to teach basic quantum mechanics to computer scientists and
  electrical engineers}.
\newblock {IEEE Trans Educ}. {2004};{47}({2}):{220--226}.
\newblock \doi{10.1109/TE.2004.825215}.

\bibitem[\protect\citeauthoryear{Singh}{{2007}}]{Singh:2006}
Singh C.
\newblock {Helping students learn quantum mechanics for quantum computing}.
\newblock AIP Conf Proc. {2007};883:{42--45}.
\newblock \doi{10.1063/1.2508687}.

\bibitem[\protect\citeauthoryear{Seegerer et~al.}{{2021}}]{Seegerer:2021}
Seegerer S, Michaeli T, Romeike R.
\newblock {Quantum computing as a topic in computer science education}.
\newblock In: {Proc. 16th Workshop Primary Secondary Comput. Educ.} 13; {2021}.
  \doi{10.1145/3481312.3481348}.

\bibitem[\protect\citeauthoryear{Rainey and Wilcox}{{2019}}]{Rainey:2019}
Rainey KD, Wilcox BR.
\newblock {Faculty survey on upper-division thermal physics content coverage}.
\newblock In: Proc. 2019 PER Conf.; {2019}. p. {494--499}.
\newblock \doi{10.1119/perc.2019.pr.Rainey}.

\bibitem[\protect\citeauthoryear{Mermin}{{2007}}]{Mermin:2007}
Mermin ND.
\newblock {Quantum computer science: An introduction}.
\newblock {Cambridge: Cambridge UP}; {2007}.

\bibitem[\protect\citeauthoryear{Nielsen and Chuang}{{2000}}]{Nielsen:2000}
Nielsen MA, Chuang IL.
\newblock {Quantum information and quantum computing}.
\newblock {Cambridge: Cambridge UP}; {2000}.

\bibitem[\protect\citeauthoryear{Meyer et~al.}{{2023}}]{Meyer:2023QST}
Meyer JC, Passante G, Wilcox BR.: The question of equity: Who has access to US
  quantum information education programs?
\newblock {In preparation for submission to \textit{Quantum Sci. Technol.}}

\bibitem[\protect\citeauthoryear{Rao and Scott}{{1981}}]{Rao:1981}
Rao JNK, Scott AJ.
\newblock {The analysis of categorical data from complex sample surveys:
  Chi-squared tests for goodness of fit and independence in two-way tables}.
\newblock {J Am Stat Assoc}. {1981};{76}(374):{221--230}.
\newblock \doi{10.1080/01621459.1981.10477633}.

\bibitem[\protect\citeauthoryear{Decady and Thomas}{{2000}}]{Decady:2000}
Decady YJ, Thomas DR.
\newblock {A simple test of association for contingency tables with multiple
  column responses}.
\newblock {Biometrics}. {2000};{56}(3):{893--896}.
\newblock \doi{10.1111/j.0006-341X.2000.00893.x}.

\bibitem[\protect\citeauthoryear{Hochberg}{{1988}}]{Hochberg:1988}
Hochberg Y.
\newblock {A sharper Bonferroni procedure for multiple tests of significance}.
\newblock {Biometrika}. {1988};{75}(4):{800--802}.
\newblock \doi{10.1093/biomet/75.4.800}.

\bibitem[\protect\citeauthoryear{Hestenes et~al.}{{1992}}]{Hestenes:1992}
Hestenes D, Wells M, Swackhamer G.
\newblock {Force concept inventory}.
\newblock {Phys Teach}. {1992};{30}:{141--158}.
\newblock \doi{10.1119/1.2343497}.

\bibitem[\protect\citeauthoryear{Wilcox et~al.}{{2015}}]{Wilcox:2015Review}
Wilcox BR, Caballero MD, Baily C, Sadaghiani H, Chasteen SV, Ryan QX, et~al.
\newblock {Development and uses of upper-division conceptual assessments}.
\newblock Phys Rev ST Phys Educ Res. {2015};{11}:{020115}.
\newblock \doi{10.1103/PhysRevSTPER.11.020115}.

\bibitem[\protect\citeauthoryear{Cataloglu and
  Robinett}{{2002}}]{Cataloglu:2002}
Cataloglu E, Robinett RW.
\newblock {Testing the development of student conceptual and visualization
  understanding in quantum mechanics through the undergraduate career}.
\newblock {Am J Phys}. {2002};{70}({3}):{238--251}.
\newblock \doi{10.1119/1.1405509}.

\bibitem[\protect\citeauthoryear{Goldhaber et~al.}{{2009}}]{Goldhaber:2009}
Goldhaber S, Pollock S, Dubson M, Beale P, Perkins K.
\newblock {Transforming upper-division quantum mechanics: Learning goals and
  assessment}.
\newblock AIP Conf Proc. {2009};{1179}:{145--148}.
\newblock \doi{10.1063/1.3266699}.

\bibitem[\protect\citeauthoryear{Wuttiprom et~al.}{{2009}}]{Wuttiprom:2009}
Wuttiprom S, Sharma MD, Johnston ID, Chitaree R, Soankwan C.
\newblock {Development and use of a conceptual survey in introductory quantum
  physics}.
\newblock {Int J Sci Educ}. {2009};{31}({5}):{631--654}.
\newblock \doi{10.1080/09500690701747226}.

\bibitem[\protect\citeauthoryear{McKagan et~al.}{{2010}}]{McKagan:2010}
McKagan SB, Perkins KK, Wieman CE.
\newblock {Design and validation of the Quantum Mechanics Conceptual Survey}.
\newblock {Phys Rev ST Phys Educ Res}. {2010};{6}:{020121}.
\newblock \doi{10.1103/PhysRevSTPER.6.020121}.

\bibitem[\protect\citeauthoryear{Singh and Zhu}{{2010}}]{Singh:2010}
Singh C, Zhu G.
\newblock {Surveying students' understanding of quantum mechanics}.
\newblock AIP Conf Proc. {2010};1289:{301--304}.
\newblock \doi{10.1063/1.3515229}.

\bibitem[\protect\citeauthoryear{Sadaghiani and
  Pollock}{{2014}}]{Sadaghiani:2014}
Sadaghiani HR, Pollock SJ.
\newblock {Quantum mechanics concept assessment: Development and validation
  study}.
\newblock {Phys Rev ST Phys Educ Res}. {2014};{11}:{010110}.
\newblock \doi{10.1103/PhysRevSTPER.11.010110}.

\bibitem[\protect\citeauthoryear{Marshman and Singh}{{2019}}]{Marshman:2019}
Marshman E, Singh C.
\newblock {Validation and administration of a conceptual survey on the
  formalism and postulates of quantum mechanics}.
\newblock {Phys Rev Phys Educ Res}. {2019};{15}:{020128}.
\newblock \doi{10.1103/PhysRevPhysEducRes.15.020128}.

\bibitem[\protect\citeauthoryear{Madsen et~al.}{{2017}}]{Madsen:2017}
Madsen A, McKagan SB, Sayre EC.
\newblock {Resource letter RBAI-1: Research-based assessment instruments in
  physics and astronomy}.
\newblock {Am J Phys}. {2017};{85}(4):{245--264}.
\newblock \doi{10.1119/1.4977416}.

\bibitem[\protect\citeauthoryear{Vignal et~al.}{{2022}}]{Vignal:2022}
Vignal M, Rainey KD, Wilcox BR, Caballero MD, Lewandowski HJ.
\newblock {Affordances of articulating assessment objectives in research-based
  assessment development}.
\newblock In: Proc. 2022 Phys. Educ. Res. Conf.; {2022}. p. {475--480}.
\newblock \doi{10.1119/perc.2022.pr.Vignal}.

\bibitem[\protect\citeauthoryear{Alpert et~al.}{{2020}}]{Alpert:2020}
Alpert CL, Edwards E, Franklin D, Freericks J.: {Key concepts for future QIS
  learners}.
\newblock 2020.
\newblock Available from:
  \url{https://files.webservices.illinois.edu/9156/keyconceptsforfutureqislearners5-20.pdf}.

\bibitem[\protect\citeauthoryear{Gerke et~al.}{{2022}}]{Gerke:2022}
Gerke F, M{\"u}ller R, Bitzenbauer P, Ubben M, Weber KA.
\newblock {Requirements for future quantum workforce - a Delphi study}.
\newblock {J Phys Conf Ser}. {2022};{2297}:{012017}.
\newblock \doi{10.1088/1742-6596/2297/1/012017}.

\bibitem[\protect\citeauthoryear{Salehi et~al.}{{2022}}]{Salehi:2022}
Salehi {\"O}, Seskir Z, Tepe {\. I}.
\newblock {A computer science-oriented approach to introduce quantum computing
  to a new audience}.
\newblock {IEEE Trans Educ}. {2022};{65}({1}):{1--8}.
\newblock \doi{10.1109/TE.2021.3078552}.

\bibitem[\protect\citeauthoryear{Angara et~al.}{{2022}}]{Angara:2022}
Angara PP, Stege U, MacLean A, M{\"u}ller HA, Markham T.
\newblock {Teaching quantum computing to high-school-aged youth: A hands-on
  approach}.
\newblock {IEEE Trans Quantum Eng}. {2022};{3}:{3100115}.
\newblock \doi{10.1109/TQE.2021.3127503}.

\bibitem[\protect\citeauthoryear{Bondani et~al.}{{2022}}]{Bondani:2022}
Bondani M, Chiofalo ML, Ercolessi E, Macchiavello C, Malgieri M, Michelini M,
  et~al.
\newblock {Introducing quantum technologies at secondary school level:
  Challenges and potential impact of an online extracurricular course}.
\newblock {Physics}. {2022};{4}({4}):{1150--1167}.
\newblock \doi{10.3390/physics4040075}.

\bibitem[\protect\citeauthoryear{Satanassi et~al.}{{2022}}]{Satanassi:2022}
Satanassi S, Ercolessi E, Levrini O.
\newblock {Designing and implementing materials on quantum computing for
  secondary school students: The case of teleportation}.
\newblock {Phys Rev Phys Educ Res}. {2022};{18}:{010122}.
\newblock \doi{10.1103/PhysRevPhysEducRes.18.010122}.

\bibitem[\protect\citeauthoryear{Kohnle and Rizzoli}{{2017}}]{Kohnle:2017}
Kohnle A, Rizzoli A.
\newblock {Interactive simulations for quantum key distribution}.
\newblock {Eur J Phys}. {2017};{38}(3):{035403}.
\newblock \doi{10.1088/1361-6404/aa62c8}.

\bibitem[\protect\citeauthoryear{DeVore and Singh}{{2020}}]{DeVore:2020}
DeVore S, Singh C.
\newblock {Interactive learning tutorial on quantum key distribution}.
\newblock {Phys Rev Phys Educ Res}. {2020};{16}:{010126}.
\newblock \doi{10.1103/PhysRevPhysEducRes.16.010126}.

\bibitem[\protect\citeauthoryear{Porter et~al.}{{2022}}]{Porter:2022}
Porter CD, Atiq Z, Fletcher E.
\newblock {Creating a modular, workforce-relevant undergraduate curriculum for
  quantum information science and engineering for all people}.
\newblock In: {Proc. 2022 Phys. Educ. Res. Conf.}; {2022}. p. {365--370}.
\newblock \doi{10.1119/perc.2022.pr.Porter}.

\bibitem[\protect\citeauthoryear{{Qubit by Qubit}}{{2023}}]{QBQ}
{Qubit by Qubit}.: Available from: \url{https://www.qubitbyqubit.org}.

\bibitem[\protect\citeauthoryear{Meyer et~al.}{{2022}}]{Meyer:2022PERC}
Meyer JC, Passante G, Pollock SJ, Wilcox BR.
\newblock {Investigating student interpretations of the difference between
  classical and quantum computers: Are quantum computers just analog classical
  computers?}
\newblock In: {Proc. 2022 Phys. Educ. Res. Conf.}; {2022}. p. {317--322}.
\newblock \doi{10.1119/perc.2022.pr.Meyer}.

\bibitem[\protect\citeauthoryear{Kushimo and Thacker}{{2022}}]{Kushimo:2022}
Kushimo T, Thacker B.
\newblock {Investigating students' strengths and difficulties in quantum
  computing}.
\newblock arXiv.
  {2022};{\href{https://arxiv.org/abs/2212.03726}{{2212.03726}}}.

\bibitem[\protect\citeauthoryear{Goorney et~al.}{2023}]{Goorney:2023}
Goorney S, Bley J, Heusler S, Sherson J.
\newblock The quantum curriculum transformation framework for the development
  of quantum information science and technology education.
\newblock arXiv. 2023;{\href{https://arxiv.org/abs/2308.10371}{{2308.10371}}}.

\bibitem[\protect\citeauthoryear{Faletic et~al.}{{2023}}]{Faletic:2023}
Faletic S, Bitzenbauer P, Bondani M, Chiofalo M, Goorney S,
  Krijtenburg-Lewerissa K, et~al.
\newblock {Contributions from pilot projects in quantum technology education as
  support action to quantum flagship}.
\newblock arXiv.
  {2023};{\href{https://arxiv.org/abs/2303.07055}{{2303.07055}}}.

\end{thebibliography}

\end{document}